\documentclass[runningheads]{llncs}

\usepackage{graphicx}
\usepackage{xcolor}
\usepackage{url}
\usepackage[hidelinks]{hyperref}
\usepackage[ruled,vlined]{algorithm2e}

\graphicspath{{fig/}}

\begin{document}

\title{Modelling Strategic Deceptive Planning \\ in Adversarial Multi-Agent Systems}

\author{Lyndon Benke\inst{1,2}
\and Michael Papasimeon\inst{1}
\and Tim Miller\inst{2}}

\authorrunning{L. Benke et al.}

\institute{Defence Science and Technology Group, Australia \and The University of Melbourne, Australia}

\maketitle

\begin{abstract}
    Deception is virtually ubiquitous in warfare, and should be a central consideration for military operations research. However, studies of agent behaviour in simulated operations have typically neglected to include explicit models of deception. This paper proposes that a computational model that approximates the human deceptive planning process would enable the authentic representation of strategic deception in multi-agent systems. The proposed deceptive planning model provides a framework for studying, explaining, and discovering deceptive behaviours, enabling the generation of novel solutions to adversarial planning problems.

    \keywords{Deception \and Multi-agent simulation \and Operations research.}
\end{abstract}

\section{Introduction}

In this position paper we propose a model of agent decision-making that explicitly incorporates the notion of deception, and discuss research challenges associated with this approach. We focus on deception concepts within the context of multi-agent-based simulations employed in operations research (OR).
Deception plays a key role during military operations, providing a means to reduce casualties and increase operational success \cite{clark_deception:_2019,daniel_strategic_1981}.
Despite its ubiquity in human conflict, deception is not explicitly considered by standard models of agent decision-making used in OR~\cite{azuma_review_2006}.

OR is the application of scientific knowledge to the study and optimisation of decision-making processes~\cite{taha_operations_2013}.
In the military domain, agent-based simulations are used to model decision-making processes in support of the effective and efficient employment of strategic and tactical capabilities \cite{tidhar_flying_1998}.
Cognitive models in OR simulations represent the decision-making processes of agents, typically implemented using techniques such as decision trees and finite-state machines.

When building cognitive models, OR analysts maintain close links with domain experts, such as military operators, to ensure the fidelity of simulated tactics and strategies.
A key requirement for these models is the \textit{explainability} of modelled behaviours, so that simulation outputs can be presented to domain experts and other clients who are not literate in agent-oriented software engineering~\cite{heinze_thinking_1998}. Computational models of deception must facilitate similar levels of explainability.

Recent work on computational deception has focused on `singular acts of deception' \cite{smith_role_2020}, where an agent attempts to maximise the deceptiveness of an action or plan according to a predefined metric \cite{kulkarni_unified_2019,liu_deceptive_2021,masters_deceptive_2017,ornik_deception_2018,root_randomized_2005}. These approaches address decision problems for which the deception itself is the goal: success is measured according to the reduced accuracy of the observer's goal recognition model.
In contrast, deception during military operations is typically strategic and extended in nature, requiring the generation of a coherent, temporally extended deceptive plan that achieves an overarching strategic goal\footnote{Operations Bodyguard and Mincemeat, Hannibal's ambush at Lake Trasimene, and even Odysseus' Trojan Horse, are examples of such strategic deceptions.}.

This paper presents a model of agent reasoning that explicitly captures a systematic deceptive planning process. It is proposed that this model facilitates the study of deceptive behaviours in OR simulations, by more accurately representing strategic military deception.
In addition, when implemented using automated planning approaches, this process may enable the discovery of novel deceptive solutions for a range of complex decision problems.

\section{Related Work} \label{sec:related-work}

The proposed deceptive planning process draws on the \textit{deception incorporation model} described by Almeshekah and Spafford in the context of computer network security \cite{almeshekah_planning_2014}, which is itself based on the more general \textit{deception cycle} described by Bell and Whaley \cite{bell_cheating_1991}. A similar process for the military domain is described by Gerwehr and Glenn \cite{gerwehr_art_2000}.
The common elements of these processes are summarised in Algorithm~\ref{algo:strategic-deception}.
These high-level models are intended for human planners, requiring a domain expert to provide a predictive model of the deception target and select appropriate deceptive techniques such as decoys and camouflage (described as the \textit{means} or \textit{instruments} of the deception).

\begin{algorithm}
\SetAlgoLined
    \Repeat{strategic goal has been achieved}
    {
        \nl Specify the \textbf{strategic goal} of the deception\;
        \nl Specify the \textbf{desired reaction} of the target to the deceptive plan\;
        \nl Decide what should be \textbf{hidden} or \textbf{shown} to achieve the desired reaction\;
        \nl \textbf{Implement} available techniques to achieve the above objectives\;
        \nl \textbf{Monitor} feedback and analyse the effect on the opponent\;
    }
    \caption{Conceptual human deceptive planning process \cite{almeshekah_planning_2014,bell_toward_2003,gerwehr_art_2000}}
    \label{algo:strategic-deception}
\end{algorithm}

Christian and Young~\cite{christian_strategic_2004} consider agent-based strategic deception in semi-cooperative scenarios, where a deceiver seeks to manipulate a target by providing planning advice with faulty information. The objective of this model is to find a plan that achieves the goals of \textit{both} the deceiver and the target. This approach is not suitable for modelling strictly adversarial scenarios, where the goals of the deceiver and target are in direct opposition.

Wagner and Arkin \cite{wagner_acting_2011} present an algorithm for acting deceptively that superficially resembles the process described in Algorithm~\ref{algo:strategic-deception}. Given a known outcome matrix and opponent model, the algorithm selects a communication that maximises the deceiver's outcome for a single action. This algorithm is therefore most similar to deception-maximising approaches \cite{kulkarni_unified_2019,liu_deceptive_2021,masters_deceptive_2017,ornik_deception_2018,root_randomized_2005}, and is not suitable for modelling strategic deception in OR simulations.

\section{Modelling Strategic Agent Deception}

\subsection{Agent Behaviour Models in Operations Research}

\begin{figure}[!htb]
	\centering
	\includegraphics[width=0.75\linewidth]{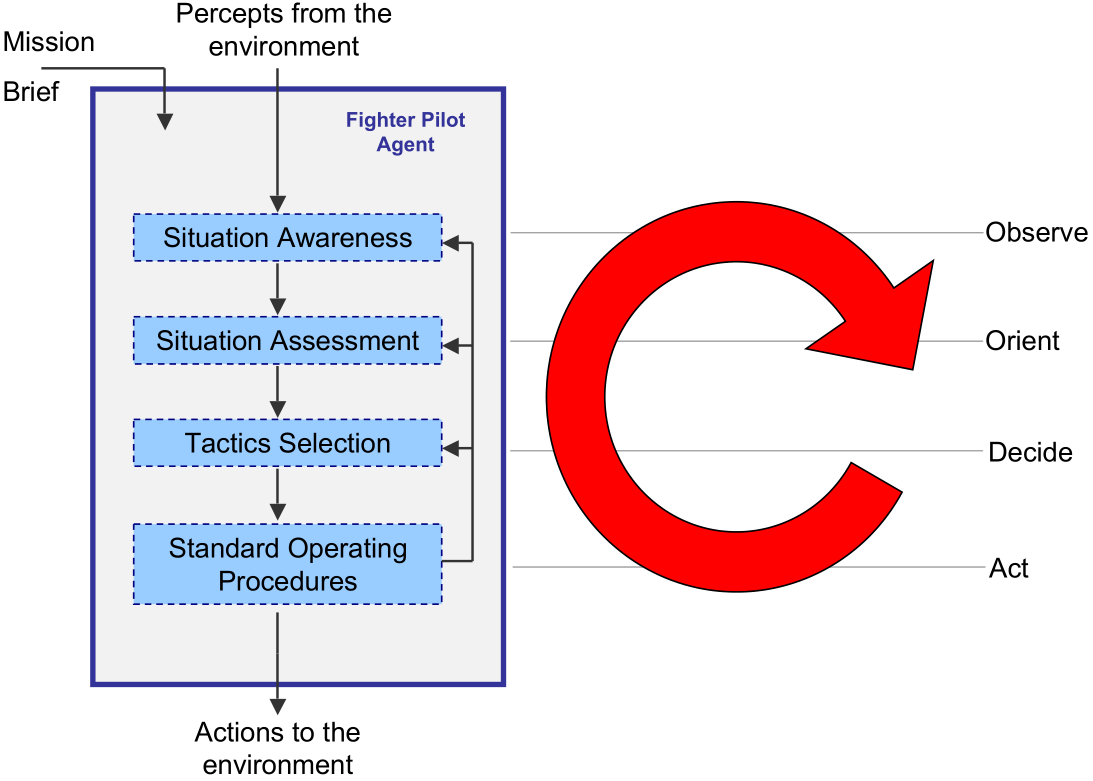}
    \caption{A variation of the OODA loop model representing the decision-making process of a fighter pilot agent (reproduced with permission) \protect\cite{papasimeon_modelling_2010}.}
    \label{fig:ooda}
\end{figure}

Agent reasoning models that support domain-friendly explanations have been used successfully in OR simulations to help explain complex behaviours to domain experts \cite{evertsz_conceptual_2017,heinze_simulating_2007}. These models, reviewed by Azuma~\textit{et~al.}~\cite{azuma_review_2006}, are typically variations on the basic \textit{observe--deliberate--act} agent control loop \cite{wooldridge_introduction_2002}.
Although originally designed for air combat, Boyd's model of fighter pilot decision-making, known as the OODA loop \cite{boyd_discourse_2018}, has become a standard decision-making model for a wide range of fields (Figure~\ref{fig:ooda}).
Previous work by Brumley \textit{et al.}~\cite{brumley_misperception_2005,brumley_orientation_2006} has observed that deception strategies may be modelled as attempts to influence an opponent's OODA loop.
Integrating an explicit deceptive planning process into Boyd's model of decision-making provides a framework for studying strategic deception that is consistent with standard OR representations of agent reasoning.

\subsection{A Strategic Deceptive Planning Model} \label{sec:deception-model}

\begin{figure}[htb]
	\centering
	\includegraphics[width=1.0\linewidth]{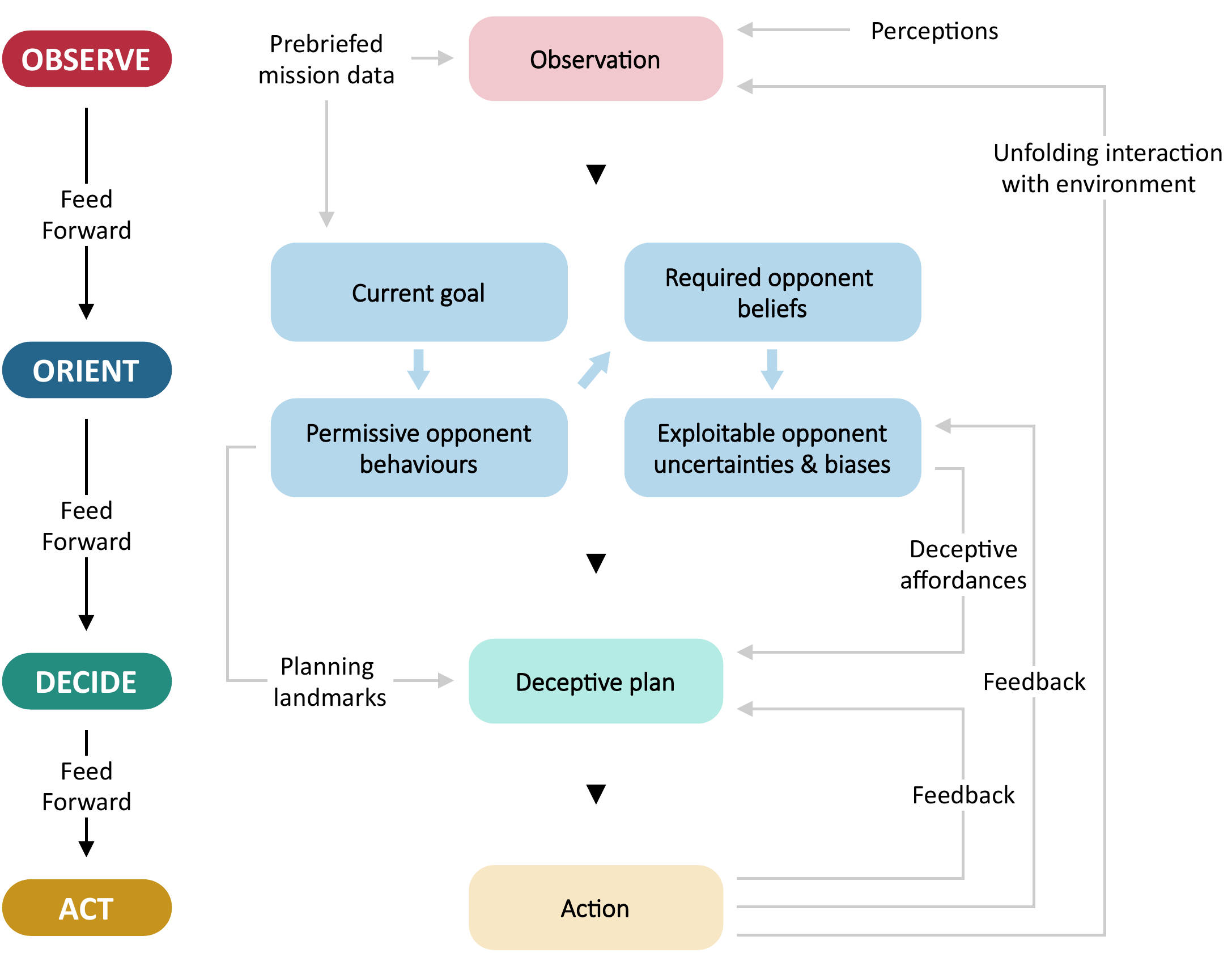}
    \caption{The proposed strategic deceptive planning model, integrating a conceptual model of human deceptive planning with the OODA control-loop model of agent reasoning. The agent identifies sets of opponent behaviours that would permit achievement of the current goal, then determines corresponding beliefs that would provoke these behaviours. The desired opponent behaviours, and any deceptive affordances that can be exploited to induce them, guide the generation of the deceptive plan.}
    \label{fig:deceptive-ooda}
\end{figure}

The proposed deceptive planning model is depicted in Figure~\ref{fig:deceptive-ooda}. This process integrates a conceptual model of human deceptive planning with the OODA control-loop model of agent reasoning. Artefacts produced during each phase of the process are fed forward to the next phase, and the cycle is repeated as new information is received from the environment.
For efficiency, the agent may choose to avoid replanning if it assesses that the unfolding interaction has not diverged significantly from the current plan. Otherwise, the agent generates an updated deceptive plan based on the new information:

\begin{enumerate}
    \item \textbf{Observe:} During the initial phase, the agent acquires data to establish situational awareness. This data includes both direct perceptions from the environment, and prebriefed information such as the mission goal and prior knowledge of the opponent. The agent processes the raw input data into more complex symbolic descriptors, updating its beliefs about the state of the environment and other agents.
    \item \textbf{Orient:} Based on the updated information, the agent assesses the situation, discovering options for achieving the current goal:
    \begin{enumerate}
        \item The agent first identifies sets of \textit{permissive opponent behaviours} that would permit achievement of the current goal, by considering a relaxed version of the planning problem in which the opponent is cooperative rather than adversarial.
        \item Next, these behaviours are used to identify corresponding \textit{required opponent beliefs} that, when present, would induce the permissive behaviours.
        \item Finally, the agent identifies perceptual and cognitive uncertainties and biases that could be exploited to enable the creation of the required opponent beliefs (described as \textit{deceptive affordances}).
    \end{enumerate}
    \item \textbf{Decide:} The agent generates a mission plan using artefacts produced during the previous phase, including the required opponent beliefs and any corresponding deceptive affordances.
    \item \textbf{Act:} Finally, the agent executes the deceptive plan. 
\end{enumerate}

As with the standard OODA loop, this model is not intended to provide a complete description of the various phases and interactions of a computational deceptive planning process. While a discussion of the details of each sub-process is beyond the scope of this paper, the remainder of this section considers challenges and opportunities that are expected to arise when implementing the proposed model in multi-agent simulation environments.

\subsection{Novel Deceptive Behaviour Discovery}

Previous studies of agent deception have generally relied on providing agents with known deceptive behaviours, either inspired by nature \cite{shim_biologically-inspired_2012}, or provided by human experts \cite{kulkarni_unified_2019,root_randomized_2005,wagner_robot_2009}. The proposed model presents an opportunity to discover deceptive solutions to planning problems without relying on existing knowledge, by enabling the integration of automated planning approaches into a strategic deceptive planning process. For example, the sub-processes identifying \textit{permissive opponent behaviours} and \textit{required opponent beliefs} lend themselves naturally to automation using diverse-planning techniques.
The automated discovery of agent behaviours is of particular interest for military OR to support the development of operational concepts for new capabilities \cite{masek_discovering_2018,park_differential_2016,ramirez_integrated_2018}.

\subsection{Deception Feasibility}
The steps in the proposed strategic deceptive planning model are subject to significant sources of uncertainty that influence whether it is possible or desirable to deceive. These include uncertainty about the true and believed states of the world (including hidden states such as opponent beliefs), and uncertainty about the ability of the agent to successfully execute the required actions.
In addition, deceptive plans often carry an additional cost for the deceiver, which may outweigh the expected benefit from the deception~\cite{wagner_acting_2011}. While a full discussion of the deception cost-risk-benefit trade-off is beyond the scope of this paper, other work has provided algorithms for deciding when it is appropriate to deceive \cite{davis_deception_2016,de_rosis_can_2003,wagner_acting_2011}.
The consideration of cost is to some extent subsumed by the presented model: if plans exist that can achieve the agent's strategic goal \textit{without} requiring a costly deception, then these plans will naturally be selected during the \textit{Orient} and \textit{Decide} phases, negating the requirement for an explicit cost-benefit analysis. This feature demonstrates an advantage of integrating the deceptive planning process into the primary agent control loop.

\subsection{Determining Exploitable Uncertainties} \label{sec:challenges-affordances}
To deceive an observer, an agent must have the capacity to influence the beliefs of the observer. The properties of the environment that enable the agent to directly create false beliefs may be considered \textit{deceptive affordances}.
\begin{definition}[Deceptive affordances] \label{def:deceptive-affordances}
    The properties of an object, agent or environment that permit deceptive actions, where a deceptive action is any action that is expected to result in the creation of a false belief for an observer.
\end{definition}
Deceptive affordances, like all affordances, are relational: the deceptive action possibilities that are available for an object, agent or environment are specific to the acting agent, and are influenced by the agent's current intentions and capabilities.
Available affordances may be known in advance, as is common in previous work on computational deception (for example, the effect of a radar jammer~\cite{davis_deception_2016}, or a distraction that prevents an agent from believing that an action has taken place~\cite{kulkarni_unified_2019}). Otherwise, deceptive affordances must be identified through experimentation in the environment. A challenge for future work will be the development of approaches to automate the discovery of exploitable uncertainties\footnote{These include both epistemic sources of uncertainty, such as sensor error and cognitive limitations, and natural statistical variability.}. A modified version of the active behaviour recognition approach described by Alford~\textit{et al.}~\cite{alford_active_2015} is proposed as a novel solution to this problem.

\section{Conclusion}

In this paper we have proposed a computational model of strategic agent deception, driven by the requirements of multi-agent-based simulation in operations research.
We described challenges and opportunities associated with this approach, particularly with respect to capturing and exploiting sources of uncertainty, and briefly discussed research directions that will attempt to address some of these challenges.
By approximating the human deceptive planning process, this model facilitates the study of strategic deception in multi-agent systems while remaining compatible with standard models of agent decision-making.
Future work will explore how automated planning techniques can be integrated with this model to enable the discovery of explainable, deceptive solutions to complex multi-agent planning problems.

\bibliographystyle{splncs04}
\bibliography{references}

\end{document}